\documentclass[10pt,twocolumn]{article}

\usepackage[utf8]{inputenc}
\usepackage{amsmath}
\usepackage{amsthm}
\usepackage{amssymb}
\usepackage{titlesec}
\usepackage{cite}
\usepackage{booktabs}
\usepackage{graphicx}
\usepackage[colorlinks=false]{hyperref}
\usepackage[format=plain,labelfont=it]{caption}
\usepackage[left=1.5cm,right=1.5cm,top=2cm,bottom=2cm]{geometry}
\usepackage{color}
\usepackage{enumitem} % Needed because itemize seems to ignore left margin otherwise
\usepackage{multirow}

\titleformat{\section}{\centering\normalfont\scshape}{\Roman{section}.}{5pt}{}
\titleformat{\subsection}{\normalfont\it}{\Alph{subsection}.}{5pt}{}
\titleformat{\subsubsection}{\normalfont\it}{\hspace{4mm}\arabic{subsubsection})}{5pt}{}

\newcommand\infoFootnote[1]{%
  \begingroup
  \renewcommand\thefootnote{}\footnote{#1}%
  \addtocounter{footnote}{-1}%
  \endgroup}

\newtheorem{thm}{Theorem}

\newtheorem{lem}[thm]{Lemma}

\newtheorem{assum}{Assumption}

\newtheorem{defn}{Definition}

\newtheorem{rem}{Remark}

\newcommand{\R}{\mathbb{R}} 

\newcommand{\ab}{\boldsymbol{a}}
\newcommand{\fb}{\boldsymbol{f}}
\newcommand{\gb}{\boldsymbol{g}}
\newcommand{\ub}{\boldsymbol{u}}
\newcommand{\xb}{\boldsymbol{x}}
\newcommand{\yb}{\boldsymbol{y}}
\newcommand{\xib}{\boldsymbol{\xi}}
\newcommand{\zerob}{\boldsymbol{0}}

\newcommand{\Ab}{\boldsymbol{A}}
\newcommand{\Bb}{\boldsymbol{B}}
\newcommand{\Cb}{\boldsymbol{C}}
\newcommand{\Db}{\boldsymbol{D}}
\newcommand{\Hb}{\boldsymbol{H}}
\newcommand{\Ib}{\boldsymbol{I}}

\newcommand{\Lb}{\boldsymbol{L}}
\newcommand{\Qb}{\boldsymbol{Q}}
\newcommand{\Rb}{\boldsymbol{R}}
\newcommand{\Ub}{\boldsymbol{U}}
\newcommand{\Wb}{\boldsymbol{W}}
\newcommand{\Yb}{\boldsymbol{Y}}
\newcommand{\Pib}{\boldsymbol{\Pi}}

\newcommand{\ybs}{\mathbf{y}}
\newcommand{\xbs}{\mathbf{x}}
\newcommand{\ubs}{\mathbf{u}}

\newcommand{\Dbc}{\boldsymbol{\mathcal{D}}}
\newcommand{\Ibc}{\boldsymbol{\mathcal{I}}}
\newcommand{\Lbc}{\boldsymbol{\mathcal{L}}}
\newcommand{\Obc}{\boldsymbol{\mathcal{O}}}
\newcommand{\Qbc}{\boldsymbol{\mathcal{Q}}}
\newcommand{\Rbc}{\boldsymbol{\mathcal{R}}}
\newcommand{\Tbc}{\boldsymbol{\mathcal{T}}}

\newcommand{\Uc}{\mathcal{U}}

\newcommand{\image}{\mathrm{image}}
\newcommand{\rank}{\mathrm{rank}}

\renewcommand{\boldsymbol}[1]{#1}
\renewcommand{\mathbf}[1]{\mathrm{#1}}

\def\tvdots{\vbox{\baselineskip=2pt \lineskiplimit=0pt \kern6pt \hbox{.}\hbox{.}\hbox{.}}}

\title{\vspace{-2mm}\bf
Extending direct data-driven predictive control towards \\systems with finite control sets
}
\author{Manuel Kl\"adtke, Moritz Schulze Darup, and Daniel E.\ Quevedo \vspace{2mm}}
\date{}

\begin{document}
\maketitle

\textbf{\textit{Abstract}.} {\bf
Although classical model predictive control with finite control sets (FCS-MPC) is quite a popular control method, particularly in the realm of power electronics systems, its direct data-driven predictive control (FCS-DPC) counterpart has received relatively limited attention. In this paper, we introduce a novel reformulation of a commonly used DPC scheme that allows for the application of a modified sphere decoding algorithm, known for its efficiency and prominence in FCS-MPC applications.
We test the reformulation on a popular electrical drive example and compare the computation times of sphere decoding FCS-DPC with an enumeration-based and a MIQP method.}
% leave no space here
\infoFootnote{M. Kl\"adtke and M. Schulze Darup are with the \href{https://rcs.mb.tu-dortmund.de/}{Control and~Cyberphysical Systems Group}, Faculty of Mechanical Engineering, TU Dortmund University, Germany. E-mails:  \href{mailto:manuel.klaedtke@tu-dortmund.de}{\{manuel.klaedtke, moritz.schulzedarup\}@tu-dortmund.de}. 
\vspace{0.5mm}}
\infoFootnote{D.~Quevedo is with Queensland University of Technology, Australia. E-mail: \href{mailto:dquevedo@ieee.org}{dquevedo@ieee.org}. 
\vspace{0.5mm}}
\infoFootnote{This paper is a \textbf{preprint} of a contribution to the 22nd European Control Conference 2024. The DOI of the original paper is \href{https://doi.org/10.23919/ECC64448.2024.10590896}{10.23919/ECC64448.2024.10590896}.}

\section{Introduction}

Model predictive control (MPC) is a widely used optimization based control strategy, popular for its ability to systematically consider system dynamics and constraints  \cite{Rawlings2017}. In power electronics applications, constraints on the control inputs often take the form of a finite control set (FCS), e.g., when they represent inverter switch positions. 
While conceptionally attractive, the FCS-MPC approach is typically associated with high computational effort, since the FCS constraints turn the associated optimal control problem into a (mixed-)integer optimization problem. To counteract this problem, different techniques have been proposed in the literature, among which the sphere decoding algorithm (SDA) stands out as the most popular choice \cite{Zafra2023}.

Recently, direct data-driven predictive control (DPC) has been introduced, which is an increasingly popular and alternative approach to MPC. The inherent flexibility and adaptability of DPC make it particularly appealing for power electronic systems, where accurate modeling is often challenging due to dynamic operation modes, uncertainties, and relaying \cite{HUANG2021}.
In its basic form, DPC utilizes linear combinations of collected trajectory data to make predictions, instead of relying on a system model (see, e.g., \cite{Coulson2019DeePC, Berberich2020, Dorfler2021}). Although exact predictions and equivalence to MPC are generally only established for linear time-invariant (LTI) systems and exact data (with some nonlinear extensions offered; see, e.g., \cite{Berberich2020}), embellishments  of DPC demonstrate encouraging results, even when these conditions are not met (e.g., \cite{Coulson2019DeePC, Dorfler2021, Coulson2019RegularizedDeePC, Berberich2020stability}). Despite the numerous extensions introduced for DPC, its use in control of systems with finite set constrained inputs, i.e., FCS-DPC, remains an unexplored area, and extending methods from FCS-MPC (like SDA) to this data-driven framework is a non-trivial task. 
\par In this paper, we address this gap by considering a DPC setup with FCS constraints and introducing an equivalent formulation that leverages the novel concept of implicit predictors introduced in \cite{KLAEDTKE2023}. This formulation readily accommodates the application of SDA, marking a first step towards addressing the computational challenges associated with FCS-DPC. We demonstrate the reduced computational burden by comparing the computation times of SDA with two other readily available methods. The structure of this paper is as follows: In Section~\ref{sec:basics}, we provide a summary of essential prerequisites for MPC, its extension to FCS-MPC, and DPC. Section~\ref{sec:FCS-DPC} elaborates on the derivation of the equivalent FCS-DPC formulation suitable for SDA. In Section~\ref{sec:Example}, we showcase the computational efficiency of this approach through simulations of an electrical drive example that is popular in the FCS-MPC literature. Finally, we conclude our work in Section~\ref{sec:Conclusion} and preview future challenges for FCS-DPC.

\section{Basics on FCS-MPC and DPC}\label{sec:basics}

In this section, we summarize some important preliminaries on FCS-MPC and DPC. For simplicity, we will consider an LTI state-space model
{
\begin{subequations}\label{eq:statespace_LTI}
\begin{align}
    \xb (k+1) &= \Ab\xb(k)+\Bb\ub(k), \quad \ub(k) \in \Uc_k\\
    \yb(k) &= \Cb\xb(k),
\end{align}
\end{subequations}
of the system with inputs $\ub\in\Uc_k\subset\R^m$, states $\xb\in\R^n$, and outputs $\yb\in\R^p$, where we assume no direct feed-through of the input to the output. Crucially, note that the input $\ub(k)$ is constrained to a finite control set $\Uc_k$, which necessitated the development of specific methods for FCS-MPC and now, similarly, FCS-DPC.

\subsection{Basics on MPC}

In classical linear MPC,  \eqref{eq:statespace_LTI} is used to predict future system behavior over a prediction horizon $N_f$. 
This model acts as an equality constraint in the associated OCP, while the control objective is chosen as a quadratic cost function and any additional inequality constraints are assumed to be linear. While there are many useful extensions to this basic setup, we will focus on the~OCP
\begin{align}
\label{eq:generalMPC}
\min_{\ub(k),\xb(k),\yb(k)} 
 &\sum_{k=0}^{N_f-1} \|\Delta\yb(k+1)\|_{\Qb}^2 +  \|\Delta\ub(k)\|_{\Rb}^2 \span \span \\
\nonumber
\text{s.t.} \quad \quad  \xb(0)&=\xb_0, \\
\nonumber
 \xb(k+1)&=\Ab\,\xb(k) + \Bb \ub(k), &&\forall k \in \{0,...,N_f-1\}, \\
 \nonumber
  \yb(k)&=\Cb\,\xb(k), &&\forall k \in \{1,...,N_f\}, \\
 \nonumber
\ub(k) & \in \Uc_k, &&\forall k \in \{0,...,N_f-1\}. 
\end{align}
Here, the cost function penalizes the squared output tracking error  $\Delta\yb(k) :=\yb(k)-\yb_\text{ref}(k)$ relative to a specified reference $\yb_\text{ref}(k)$ and the squared change in control action $\Delta\ub(k):=\ub(k)-\ub(k-1)$, where $\Qb\in \R^{p\times p}, \Rb\in \R^{m\times m}$ are symmetric positive semi-definite and positive definite weighing matrices, respectively. Note that $\xb_0$ refers to the most recently measured state in closed-loop and (with some abuse of notation) $k$ is used as a relative time index for predictions in \eqref{eq:generalMPC}, while it may refer to an absolute time index at other points in this paper. To close the loop, \eqref{eq:generalMPC} is solved at every time-step and only the first element $\ub^\ast(0)$ of the optimal control sequence is applied to the system. In the most simple setup, the control sets $\Uc_k$ are polyhedral, making the OCP \eqref{eq:generalMPC} not only convex but also allow for it to be cast into a quadratic program (QP), which can be solved efficiently. In contrast, for FCS-DPC, $\Uc_k$ is totally disconnected resulting in non-convex input constraints (neglecting the trivial case, where $\Uc_k$ are singletons).

\subsection{Solving FCS-MPC via sphere decoding}\label{sec:FCS-MPC}

Assuming that the elements of $\Uc_k$ are   integers (this is without loss of generality, since they can simply be mapped to integers otherwise), the OCP \eqref{eq:generalMPC} can be cast into a mixed-integer QP (MIQP), which are generally NP-hard. Perhaps the most naive way of solving the FCS-MPC problem would be to enumerate all elements of $\Uc := \Uc_0 \times \hdots \times \Uc_{N_f-1}$, compute the resulting predicted state and output trajectories via \eqref{eq:statespace_LTI}, and compare the associated costs. However, as the cardinality of $\Uc$ scales drastically (i.e., exponentially for $|\Uc_k| = \text{const.}$) with $N_f$, this approach is typically intractable for real-time computation in closed-loop, except for very small $N_f$, where $N_f = 1$ seems to be a popular choice in the power electronics community \cite{Muller2011}. Thankfully, more sophisticated methods exist for (mixed-)integer programming, which often use a branch-and-bound \cite{BranchAndBound1960} procedure to determine suboptimality for large subsets of $\Uc$ without needing to explicitly evaluate the associated cost of each element. Specifically in FCS-MPC, a modified version of the sphere decoding algorithm (SDA) \cite[Alg.~1]{Geyer2014} has become popular, which can indeed be classified as a branch-and-bound method as illustrated, e.g., in \cite[Fig.~3]{Zafra2023}. In general, sphere decoding is an efficient method for closest (Euclidean distance) point search in lattices. In the following, we briefly summarize how this version of SDA can be applied to FCS-MPC, where we assume a time-invariant control set, i.e., $\Uc_0 = \hdots = \Uc_{Nf-1}$, for simplicity.
First, note that \eqref{eq:statespace_LTI} is a one-step prediction model that can be consecutively applied to yield the multi-step predictor $\ybs_f = \Obc \xb_0 + \Tbc \ubs_f$ with
$$
    \Obc := 
    \begin{pmatrix}
        \Cb \\
        \Cb\Ab \\ 
        \vdots \\
        \Cb\Ab^{N_f-1}
    \end{pmatrix}\,\,\text{and}\,\,
        \Tbc := 
    \begin{pmatrix}
     \zerob  & &  & \!\zerob\\
    \Cb \Bb   &  \!\ddots &  & \\
    \vdots &  \!\ddots &  \!\ddots &  \\
    \Cb \Ab^{N_f-2} \Bb  & \!\dots & \!\!\Cb \Bb & \!\zerob
    \end{pmatrix}
$$
and $\ubs_f \in \R^{m N_f}, \ybs_f \in \R^{p N_f}$ being vectors containing the (stacked) predicted input and output sequences, respectively. Next, we stack the sequences $\yb_\text{ref}(k)$ and $\Delta u(k)$ similarly, where we find $\Delta \ubs_f = \Ibc \ubs_f - \Lbc\ub(-1)$ with
$$
    \Ibc := \begin{pmatrix}
        \Ib_m & \zerob & \hdots & \zerob \\
        -\Ib_m & \Ib_m & & \\
         & \ddots & \ddots  \\
         \zerob & & - \Ib_m & \Ib_m
    \end{pmatrix} 
    \,\,\text{and}\,\,
    \Lbc := 
    \begin{pmatrix}
     \Ib_m \\ \zerob \\ \vdots \\ \zerob
    \end{pmatrix},
$$
and the previous time-step input $\ub(-1)$ will act as a parameter, similarly to $\xb_0$. Using these, the cost function of the OCP \eqref{eq:generalMPC} can be reformulated in standard quadratic form $\frac{1}{2} \ubs_f^\top \Hb \ubs_f + \fb^\top \ubs_f + c$, where $c$ is a constant term with respect to $\ubs_f$ (i.e., irrelevant for the optimizer) and the remaining parameters are given by
\begin{align*}
   \Hb &:= 2 \left(\Tbc^\top \Qbc\Tbc + \Ibc^\top\Rbc\Ibc \right), \\
    \fb &:= 2 \left(\left(\Obc\xb_0-\ybs_\text{ref}\right)^\top\Qbc\Tbc+\left(\Lbc \ub(-1)\right)^\top \Rbc \Ibc\right)^\top,
\end{align*}
and $\Qbc:=\text{diag}(\Qb, \hdots, \Qb), \Rbc:=\text{diag}(\Rb, \hdots, \Rb)$. This procedure effectively eliminates the predicted state sequence $\xbs_f$ and output sequence $\ybs_f$ from the OCP, leaving only the FCS-constrained input sequence  $\ubs_f$ as an optimization variable. If we were to neglect the FCS constraint $\ubs_f\in\Uc$ for a moment, the unconstrained optimum would be given by $\ubs_{f,\text{unc}}^\ast = -\Hb^{-1}\fb$ and the cost function can be equivalently stated in terms of a squared distance to this unconstrained optimum, i.e., $\frac{1}{2}\big(\ubs_f-\ubs_{f,\text{unc}}^\ast\big)^\top\Hb\big(\ubs_f-\ubs_{f,\text{unc}}^\ast\big)$ by completing the square, where irrelevant constant terms have already been discarded. This distance is not quite the Euclidean distance needed for sphere decoding due to the Hessian $\Hb$ acting as a weighing matrix. However, since $\Hb$ is positive definite (due to $\Rb$ being chosen positive definite) and symmetric, we can use the Cholesky decomposition of $\Hb^{-1} = \Lb^{-1} \Lb^{-\top}$ (where $\Lb$ is a lower triangular and positive definite matrix such that $\Lb^\top \Lb = \Hb$) to define a coordinate transformation $\tilde \ubs_f := \Lb \ubs_f$ (and similarly for $\tilde \ubs_{f,\text{unc}}$). Using $\Lb$, the optimal input sequence $\ubs_f^\ast$ solving the FCS-MPC problem can finally be found by a (truncated) integer least-squares problem
\begin{equation}\label{eq:FCS-MPC}
    \ubs_f^\ast = \arg\min_{\ub_{f}\in \Uc} \|\Lb \ubs_f -  \tilde \ubs_{f,\text{unc}}\|_2^2,
\end{equation}
to which the modified SDA can be applied. Unfortunately, a full explanation of this algorithm and discussion of its variants is outside the scope of this paper, so we reference the excellent explanations in \cite{Geyer2014, Zafra2023}, instead. However, note that, crucially, if the FCS-DPC problem can be stated similarly to \eqref{eq:FCS-MPC}, the same algorithm can be applied for its solution. Hence, this reformulation will be the focus of this paper and treated in Section~\ref{sec:FCS-DPC}, after introducing some basics on DPC in the following section.

\subsection{Basics on regularized DPC}\label{sec:DPC_basics}

In contrast to the linear MPC approach, which typically relies on an LTI model \eqref{eq:statespace_LTI}, DPC utilizes predictions derived from previously collected trajectory data $(\ubs^{(1)}, \ybs^{(1)}), \hdots, (\ubs^{(\ell)}, \ybs^{(\ell)})$. These predictions are constructed using linear combinations
$$
    \begin{pmatrix}
        \ubs_\text{pred}\\
        \ybs_\text{pred}
    \end{pmatrix}
    =
    \begin{pmatrix}
        \ubs^{(1)}\\
        \ybs^{(1)}
    \end{pmatrix} a_1 + \hdots + 
    \begin{pmatrix}
        \ubs^{(\ell)}\\
        \ybs^{(\ell)}
    \end{pmatrix} a_\ell
    = 
    \Dbc \ab.
$$
Here, the dimensions of the data matrix $\Dbc\in\R^{L(m+p)\times\ell}$ and the generator vector $\ab\in\R^\ell$ are determined by the length $L$ of recorded (as well as predicted) trajectories and the number $\ell$ of data trajectories used for predictions. This approach is grounded in a result applicable to LTI systems, where, assuming that $L$ exceeds the system's lag, $\image(\Dbc)$ is equivalent to the set of all possible system trajectories if and only if \cite{Markovsky2020} 
\begin{equation}\label{eq:GPE}
    \rank(\Dbc)=L m + n. 
\end{equation}
It is important to note that the condition \eqref{eq:GPE} not only establishes a minimum rank requirement for capturing the whole system behavior but also signifies the maximum rank that the data matrix $\Dbc$ can achieve for exact data.
In situations where individual trajectories $(\ubs^{(i)}, \ybs^{(i)})$ correspond to time-shifted sections of a single long trajectory, a well-known sufficient condition for satisfying \eqref{eq:GPE} is provided by Willems' fundamental lemma \cite{WILLEMS2005}. To incorporate the current initial condition of the system as the starting point for predicted trajectories, the predicted I/O-sequence is typically divided into two segments: a past section $(\ubs_p, \ybs_p)$ and a future section $(\ubs_f, \ybs_f)$ with $N_p$ and $N_f$ time-steps, respectively, yielding
$$
    \begin{pmatrix}
        \ubs_\text{pred}\\
        \ybs_\text{pred}
    \end{pmatrix}
    = 
    \begin{pmatrix}
        \ubs_p\\
        \ubs_f\\
        \ybs_p \\
        \ybs_f
    \end{pmatrix}
    =
    \begin{pmatrix}
        \Ub_p\\
        \Ub_f\\
        \Yb_p \\
        \Yb_f
    \end{pmatrix}\ab
    =
    \Dbc \ab. 
$$
The past section of a predicted trajectory is then constrained to match the I/O-data $\xib$ recorded in the most recent $N_p$ time-steps during closed-loop operation. In other words, the equality constraints
$$
    \xib= 
    \begin{pmatrix}
        \ubs_p \\ \ybs_p
    \end{pmatrix}
    =
    \begin{pmatrix}
        \Ub_p \\ \Yb_p
    \end{pmatrix}\ab
    = \Wb_p \ab
$$
ensure that any predicted trajectory in \eqref{eq:DPC} begins with the most recently observed behavior of the system. In this context, the past trajectory $\xib$ can also be interpreted as the state of a (typically non-minimal) state-space realization of the system, and specifies the system's initial condition when $N_p$ is selected to be equal to or greater than its lag \cite{Markovsky2008}.
\begin{rem}
    Due to how we set up the MPC problem with no feed-through term ($\Db = 0$) in \eqref{eq:generalMPC}, we likewise modify the DPC setup to predict 
    $$
        \ubs_f = 
        \begin{pmatrix}
            \ub(0) \\
            \vdots \\
            \ub(N_f-1)
        \end{pmatrix}, \quad 
        \ybs_f = 
        \begin{pmatrix}
            \yb(1) \\
            \vdots \\
            \yb(N_f)
        \end{pmatrix}. 
    $$
    Note that this does not conflict with the theory supporting direct data-driven predictions as it is equal to first extending the prediction horizon by one step (i.e., predicting system behavior in the (relative) time-steps $k \in\{0, ..., N_f$\}) and then neglecting $\ub(N_f)$ and $\yb(0)$ in the OCP formulation.
\end{rem}

Crucially, it has been shown in \cite{Coulson2019DeePC} that DPC given by\begin{subequations}
\label{eq:DPC}
\begin{align}
\min_{\ubs_f,\ybs_f,\ab} 
 \|\ybs_f-\ybs_\text{ref}\|_{\Qbc}^2 &+  \|\Delta\ubs_f\|_{\Rbc}^2+h(\ab) \label{eq:DPCcost} \\
\text{s.t.} \quad \quad  \begin{pmatrix}
    \xib \\ \ubs_f \\ \ybs_f 
\end{pmatrix} &= \begin{pmatrix}
    \Wb_p \\ \Ub_f \\ \Yb_f 
\end{pmatrix}\ab, \label{eq:DPCeqConstr}\\
\ubs_f &\in \Uc  \label{eq:DPC_FCSConstr}
\end{align}
\end{subequations}
based on exact data generated by an LTI system and without regularization ($h(\ab) = 0$) is equivalent to the MPC in \eqref{eq:generalMPC}, since \eqref{eq:DPCeqConstr} is an exact (image) representation of the system behavior. Note that this result holds for general constraint sets $\Uc$, and thus also for the FCS. While this basic approach works for exact measurement data generated by an LTI system, it generally fails whenever more realistic settings with nonlinearities, noise, or disturbances are considered. Generally speaking, their presence in the data causes the space spanned by $\image(\Dbc)$ not only to be skewed, but also typically increases its dimension. The latter allows for unreasonable predictions, which are inevitably exploited with respect to the cost function while solving the OCP. In the most extreme case, $\Dbc$ has full row rank, which renders the constraint \eqref{eq:DPCeqConstr} meaningless, since there exists a value $\ab$ solving the equation for any arbitrary left-hand side. For a more detailed explanation of this effect, see, e.g., \cite[Sec.~6]{Mattsson2021}. The currently accepted remedy for this effect (first proposed in \cite{Coulson2019DeePC}) is the addition of a regularization term $h(\ab)$ to the cost function, which (ideally) restores meaning to \eqref{eq:DPCeqConstr} and renders the resulting cost function a trade-off between optimism (given by the original control objective cost function) and realism (given by the regularization). There exist several useful interpretations of this regularization effect for different choices of $h(\ab)$. In the present work, we will focus on a quadratic regularization $h(\ab) = \lambda_a \|\ab\|_2^2$ and its variation $h(\ab) = \lambda_a \|(\Ib-\Pib)\ab\|_2^2$ utilizing the projection matrix
$$
    \Pib := \begin{pmatrix}
        \Wb_p \\
        \Ub_f
    \end{pmatrix}^\top\left(
    \begin{pmatrix}
        \Wb_p \\
        \Ub_f
    \end{pmatrix}
    \begin{pmatrix}
        \Wb_p \\
        \Ub_f
    \end{pmatrix}^\top
    \right)^{-1}
    \begin{pmatrix}
        \Wb_p \\
        \Ub_f
    \end{pmatrix}
$$
introduced in \cite{Dorfler2021}, since they are popular choices if the true system dynamics are assumed to exhibit (close to) LTI behavior. Among existing interpretations, we reference \cite{Dorfler2021, Coulson2019RegularizedDeePC} for sake of completeness but focus on the concept of implicit predictors introduced in \cite{KLAEDTKE2023}. The latter will be utilized to derive equivalent OCP formulations in the following.

\section{Applying sphere decoding to FCS-DPC}\label{sec:FCS-DPC}

As noted in Section~\ref{sec:FCS-MPC}, the focus of this main section lies in deriving an equivalent OCP for the DPC problem \eqref{eq:DPC} with FCS-constraint, to which the modified sphere decomposition algorithm can be applied. In the following, we recall the novel concept of implicit predictors introduced in \cite{KLAEDTKE2023}, derive an implicit predictor for the DPC problem \eqref{eq:DPC} with FCS-constraints, and apply it as an explicit constraint to derive an OCP formulation, to which the SDA is applicable.

\subsection{An implicit predictor for FCS-DPC}\label{sec:FCS-DPC_predictor}

Contrary to MPC, an output predictor in the sense of an input-(state)-output mapping is not explicitly present as a constraint in \eqref{eq:DPC}. Nevertheless, we can characterize such a mapping implicitly in the following way.
\begin{defn}[\cite{KLAEDTKE2023}]\label{def:implicit_predictor}
    We call $\hat\ybs(\xb_0,\ubs_f)$ an \textit{implicit predictor} for an OCP if including the constraint $\ybs_f = \hat\ybs(\xb_0, \ubs_f)$ does not alter the (set of) minimizers $(\ubs_f^\ast, \ybs_f^\ast)$ and optimal value.  
\end{defn}
In the case of DPC, the past I/O-sequence $\xib$ takes the role of a state $\xb_0$. Conceptually, the implicit predictor serves as a valuable tool for characterizing the predictive behavior of DPC, as it inherently aligns with this behavior by definition. In this context, its primary advantage can be viewed as a descriptive mapping (though we will employ it as a prescriptive mapping in Section~\ref{sec:FCS-DPC}), shedding light on the somewhat opaque predictive nature of DPC, as extensively discussed in \cite{KLAEDTKE2023}. 
Similarly to \cite{Mattsson2021, KLAEDTKE2023}, we make the following assumption to avoid special cases and to simplify the analysis.
\begin{assum}\label{assum:fullRank}
    The data matrix $\Dbc$ has full row rank.
\end{assum}
While this is the extreme case mentioned in \ref{sec:DPC_basics}, it is also a reasonable assumption (as noted, e.g., in \cite[Sec.~3]{Mattsson2021} and \cite[Sec.~3]{KLAEDTKE2023}) in the presence of measurement noise, as long as enough measurement data is available to make $\Dbc$ at least square, i.e., $\ell \geq (m+p)(N_p+N_f)$. Crucially, note that the specification (including choice of $h(\ab)$) of the DPC problem \eqref{eq:DPC} and Assumption~\ref{assum:fullRank} are sufficient to determine the predictive behavior in the sense of Definition~\ref{def:implicit_predictor}. That is, we do not make any additional assumptions on the system dynamics or type of noise present in the data. This is because an implicit predictor as in Definition~\ref{def:implicit_predictor} only characterizes the predictive behavior that the DPC scheme attributes to the data based on $\Dbc$, and does not necessarily align with the true system behavior. Essentially, if \eqref{eq:DPC} is poorly tuned or otherwise generally unfit to properly predict the true system behavior, the use of implicit predictors will not change that, but their analysis might shine a light on this problem. To prepare for the derivation of the implicit predictor in Theorem~\ref{thm:implicitPredictor2FCS}, we also recall the following Lemma (slightly adjusted to the cost function of \eqref{eq:DPC}).
\begin{lem}[\cite{KLAEDTKE2023}]\label{lem:eliminate_a}
    Under Assumption~\ref{assum:fullRank}, the regularized DPC problem \eqref{eq:DPC} is equivalent to   
    \begin{align}\label{eq:regDPCouter}
    \min_{\ubs_f,\ybs_f} &
     \|\ybs_f-\ybs_\text{ref}\|_{\Qbc}^2 +  \|\Delta\ubs_f\|_{\Rbc}^2 + h^\ast(\xib, \ubs_f, \ybs_f) \\ 
     & \text{s.t.}\quad  \eqref{eq:DPC_FCSConstr} \nonumber
    \end{align}
    with unique
    \begin{equation}\label{eq:regDPCinner}
    h^\ast(\xib, \ubs_f, \ybs_f) := \min_{\ab}\:\:   h(\ab) \quad 
    \text{s.t.}\quad \eqref{eq:DPCeqConstr}.
    \end{equation}
\end{lem}
This lemma effectively allows eliminating $\ab$ and \eqref{eq:DPCeqConstr} from the OCP, since (due to Assumption~\ref{assum:fullRank}) their only effect is fully captured by $h^\ast(\xib, \ubs_f, \ybs_f)$. Furthermore, analytic solutions to \eqref{eq:regDPCinner} are derived in \cite{KLAEDTKE2023} as 
\begin{align}
     h^\ast(\xib, \ubs_f, \ybs_f) &= \lambda_a \|\ybs_f-\hat\ybs_\text{SPC}(\xib, \ubs_f)\|_{\Qbc_\text{reg}}^2, \label{eq:proj2NormCost} \\
     h^\ast(\xib, \ubs_f, \ybs_f) &= \lambda_a \|\ybs_f-\hat\ybs_\text{SPC}(\xib, \ubs_f)\|_{\Qbc_\text{reg}}^2 \label{eq:2NormCost}\\
    &\quad + \lambda_a \|\ubs_f-\Ub_f \Wb_p^+\xib\|_{\Rbc_\text{reg}}^2 + \lambda_a\|\xib\|_{\left(\Wb_p\Wb_p^\top\right)^{-1}}^2 \nonumber
\end{align}
for $h(\ab)=\lambda_a\|(\Ib-\Pib)\ab\|_2^2$ and $h(\ab)=\lambda_a\|\ab\|_2^2$, respectively. Here, we have simplified the expression in \cite[Eq.~(14)]{KLAEDTKE2023} via block matrix inversion formulas and introduced
\begin{align*}
    \Qbc_\text{reg} &:=\left(\Yb_f\left(\Ib-\Pib\right)\Yb_f^\top\right)^{-1} \\
    \Rbc_\text{reg} &:=\left(\Ub_f\left(\Ib-\Wb_p^\top\left(\Wb_p\Wb_p^\top\right)^{-1}\Wb_p\right)\Ub_f^\top\right)^{-1}
\end{align*}
to shorten the notation. Furthermore, the mapping
$$
    \hat\ybs_\text{SPC}(\xib,\ubs_f):= \Yb_f \begin{pmatrix}
    \Wb_p \\ \Ub_f
\end{pmatrix}^+\begin{pmatrix}
    \xib \\ \ubs_f
\end{pmatrix}
$$
refers to the Subspace Predictive Control (SPC) \cite{FAVOREEL1999} predictor, where we will sometimes omit the arguments for brevity. The connection of SPC and DPC (with deterministic LTI data or quadratic regularization) is thoroughly discussed in, e.g., \cite{Fiedler2021, Dorfler2021, Mattsson2021, KLAEDTKE2023}, thus we defer to these sources for detailed discussions. Given these preliminaries, an implicit predictor for the FCS-DPC problem can be specified as follows.
\begin{thm}\label{thm:implicitPredictor2FCS}
    Consider the FCS-constrained DPC problem~\eqref{eq:DPC} with regularizer $h(\ab)=\lambda_a\|\ab\|_2^2$ or $h(\ab)={\lambda_a\|(\Ib-\Pib)\ab\|_2^2}$. 
    Under Assumption~\ref{assum:fullRank},
\begin{align}
 \label{eq:FCS-DPC_predictor}
        \hat\ybs_\text{DPC}(\xib, \ubs_f)=&
        \left(\lambda_a \Qbc_\text{reg}+\Qbc\right)^{-1} \lambda_a \Qbc_\text{reg} \hat\ybs_\text{SPC}(\xib,\ubs_f)\\
        &\;+\left(\lambda_a \Qbc_\text{reg}+\Qbc\right)^{-1} \Qbc \ybs_\text{ref} \nonumber
\end{align}
    is an implicit predictor for this problem.
\end{thm}
\begin{proof}
    The implicit predictor \eqref{eq:FCS-DPC_predictor} is the minimizer 
    $$
        \hat\ybs_\text{DPC}(\xib, \ubs_f)=\arg\min_{\ybs_f} 
     \|\ybs_f-\ybs_\text{ref}\|_{\Qbc}^2  + h^\ast(\xib, \ubs_f, \ybs_f) 
    $$
    to an inner optimization problem for \eqref{eq:regDPCouter}, where not only $\xib$ but also $\ubs_f$ act as parameters. Crucially, since $\ubs_f$ is not an optimization variable in this problem, the FCS-constraint \eqref{eq:DPC_FCSConstr} and cost term $\|\Delta\ubs_f\|_\Rbc^2$ are irrelevant here and have thus been dropped. Furthermore, note that the only difference between $h^\ast(\xib, \ubs_f, \ybs_f)$ for the two considered regularizers are terms independent of $\ybs_f$ and thus equally negligible. Since dropping any further terms independent of $\ybs_f$ does not change the minimizer, we can simplify the problem as
    \begin{align*}
        &\arg\min_{\ybs_f}  
     \|\ybs_f-\ybs_\text{ref}\|_\Qbc^2+\lambda_a \|\ybs_f-\hat\ybs_\text{SPC}\|_{\Qbc_\text{reg}}^2 \\
     =& \arg\min_{\ybs_f} \ybs_f^\top \left(\lambda_a\Qbc_\text{reg}\!+\!\Qbc\right)\ybs_f\! -\! 2 \left(\hat\ybs_\text{SPC}^\top \lambda_a\Qbc_\text{reg}\!+\!\ybs_\text{ref}^\top \Qbc\right) \ybs_f,
    \end{align*}
    which yields an unconstrained quadratic minimization problem with the minimizer given by \eqref{eq:FCS-DPC_predictor}. Now, since $\hat\ybs_\text{DPC}(\xib, \ubs_f)$ is the parametric minimizer for any $(\xib,\ubs_f)$, the minimizers $(\ubs_f^\ast, \ybs_f^\ast)$ to the regularized DPC problem must naturally satisfy the relation $\ybs_f^\ast = \hat\ybs_\text{DPC}(\xib, \ubs_f^\ast)$ for any $\xib$. Hence, including the equality constraint $\ybs_f = \hat\ybs_\text{DPC}(\xib, \ubs_f)$ with the regularized DPC problem does not change its optimal value or minimizers, making $\hat\ybs_\text{DPC}(\xib, \ubs_f)$ as given by \eqref{eq:FCS-DPC_predictor} an implicit predictor of this OCP.
\end{proof}
Crucially, the implicit predictor characterized in Theorem~\ref{thm:implicitPredictor2FCS} is  not affected by the FCS constraint, since input constraints do not affect the predictive behavior in the sense of Definition~\ref{def:implicit_predictor}, as already noted in \cite[Sec.~III.C]{KLAEDTKE2023}. Therefore, this implicit predictor is very similar to those characterized in \cite[Thm.~3, 4]{KLAEDTKE2023}, differing only by a linear term in the cost function stemming from output reference tracking, which results in an additional constant term in the implicit predictor given by \eqref{eq:FCS-DPC_predictor}. The predictive behavior is therefore not linear but affine, as already noted in \cite[Rem.~2]{KLAEDTKE2023}. Similarly to the MPC multi-step predictor, we partition the affine implicit predictor $\hat\ybs_\text{DPC}(\xib, \ubs_f)= \Obc_\text{DPC} \xib + \Tbc_\text{DPC} \ubs_f + \gb_\text{DPC}$ and the linear SPC predictor $\hat\ybs_\text{SPC}(\xib,\ubs_f)=\Obc_\text{SPC} \xib + \Tbc_\text{SPC}\ubs_f$ to isolate the effect of the ``state'' $\xib$, input sequence $\ubs_f$ and constant term. Furthermore, since $h^\ast(\xib, \ubs_f, \ybs_f)$ penalizes the difference between $\ybs_f$ and $\hat\ybs_\text{SPC}(\xib, \ubs_f)$ as apparent from \eqref{eq:proj2NormCost} and \eqref{eq:2NormCost}, we introduce $\Delta\Obc := \Obc_\text{DPC}-\Obc_\text{SPC}$ and $\Delta\Tbc := \Tbc_\text{DPC}-\Tbc_\text{SPC}$ for concise notation.

\subsection{Reformulating FCS-DPC for sphere decoding}

As per Lemma~\ref{lem:eliminate_a}, we can effectively eliminate $\ab$ and \eqref{eq:DPCeqConstr} by considering \eqref{eq:regDPCouter} instead. Furthermore, adding $\ybs_f = \hat\ybs_\text{DPC}(\xib, \ubs_f)$ as an equality constraint does not change the solution of the OCP as per Definition~\ref{def:implicit_predictor}. Therefore, instead of the original FCS-DPC problem \eqref{eq:DPC}, we can consider the equivalent OCP
\begin{subequations}\label{eq:FCS-DPC_indirect}
    \begin{align}
    \min_{\ubs_f,\ybs_f} 
     \|\ybs_f-\ybs_\text{ref}\|_{\Qbc}^2 &+  \|\Delta\ubs_f\|_{\Rbc}^2 + h^\ast(\xib, \ubs_f, \ybs_f) \\ 
      \text{s.t.}\quad  
     \ybs_f &= \Obc_\text{DPC} \xib + \Tbc_\text{DPC} \ubs_f + \gb_\text{DPC}  \label{eq:explicitPredictor}\\ 
     \ubs_f & \in \Uc. 
\end{align}
\end{subequations}
\begin{rem}\label{rem:directiVSindirect}
We acknowledge that this essentially turns the direct data-driven approach of \eqref{eq:DPC} into an indirect one by explicitly enforcing an estimated model given by $\hat\ybs_\text{DPC}(\xib, \ubs_f)$. However, since the enforced system model is the implicit predictor of the direct approach and the objective function is kept equally intact by maintaining the effect of the regularization via $h^\ast(\xib, \ubs_f, \ybs_f)$, both are provably equivalent. Hence, any advantages or disadvantages in closed-loop performance that the DPC approach may have compared to traditional indirect approaches are equally maintained by \eqref{eq:FCS-DPC_indirect}. Furthermore, \eqref{eq:FCS-DPC_indirect} has the advantage that the optimization variables remain independent of the number $\ell$ of columns in $\Dbc$, since $\ab \in \R^\ell$ has been eliminated via Lemma~\ref{lem:eliminate_a}.
\end{rem}
Focusing on the case $h(\ab) = \|(\Ib-\Pib)\ab\|_2^2$ in \eqref{eq:proj2NormCost} for a moment, we proceed similarly to Section~\ref{sec:FCS-MPC} by eliminating $\ybs_f$ via \eqref{eq:explicitPredictor} and reformulate the cost function in quadratic form $\frac{1}{2} \ubs_f^\top \check\Hb \ubs_f + \check\fb^\top \ubs_f$ with
\begin{align*}
    \check\Hb &:= 2 \; \Tbc_\text{DPC}^\top \Qbc\Tbc_\text{DPC} + 2 \Ibc^\top\Rbc\Ibc  + 2 \lambda_a \Delta \Tbc^\top \Qbc_\text{reg}\Delta \Tbc , \\
    \check\fb &:= 2 \Tbc_\text{DPC}^\top \Qbc\left(\Obc_\text{DPC}\xib+\gb_\text{DPC}-\ybs_\text{ref}\right)+ \Ibc^\top \Rbc \Lbc \ub(-1)  \\
    & \qquad \quad +\lambda_a \Delta \Tbc^\top \Qbc_\text{reg} \left(\Delta \Obc\xib+\gb_\text{DPC}\right),
\end{align*}
where we already dropped all cost terms that are constant with respect to $\ubs_f$. For the case $h(\ab) = \|\ab\|_2^2$ in \eqref{eq:2NormCost}, only the terms $2 \lambda_a \Rbc_\text{reg}$ and $-2 \lambda_a \Rbc_\text{reg}\Ub_f\Wb_p^+\xib$ need to be added to $\check\Hb$ and $\check\fb$, respectively. Given these specifications, the remaining process is, again, similar to Section~\ref{sec:FCS-MPC}. We use the Cholesky decomposition of $\check\Hb^{-1} = \check\Lb^{-1} \check\Lb^{-\top}$ to define the coordinate transformation $\check \ubs_f := \check \Lb \ubs_f$ and finally state the optimal control sequence 
\begin{equation}\label{eq:FCS-DPC}
    \ubs_f^\ast =\arg\min_{\ub_{f}\in \Uc} \|\check\Lb \ubs_f -  \check \ubs_{f,\text{unc}}\|_2^2,
\end{equation}
as the solution of a truncated least-squares problem, to which the modified SDA \cite[Alg.~1]{Geyer2014} can be applied. Unsurprisingly, the transformed unconstrained optimum $\check \ubs_{f,\text{unc}} = -\check\Lb\check\Hb^{-1}\check\fb$ is defined similarly as for FCS-MPC. Note that, while the implicit predictor \eqref{eq:FCS-DPC_predictor} may be time-variant due to changing references $\ybs_\text{ref}$, this does not require a re-computation of $\check\Hb$ or $\check\Lb$ during closed-loop operation. Instead, $\gb_\text{DPC}$ only affects $\check\fb$, which already needs re-computation in every time-step due to its dependence on $\xib, \ub(-1)$, and $\ybs_\text{ref}$.

\section{Example}\label{sec:Example}

To illustrate this FCS-DPC approach, we consider a three-level three-phase neutral point clamped voltage source inverter driving an induction motor with fixed neutral point example that is popular in the FCS-MPC literature, e.g., \cite{Geyer2013, Geyer2014, Stellato2017, Zafra2023}. A detailed description of the setup and its state-space model can be found in \cite{Geyer2013, Geyer2014, Stellato2017}. Some example system parameters are provided in \cite{Geyer2013} and we used the code provided with \cite{Stellato2018} to help set up our simulation. Note that this example includes an additional switching constraint $\|\Delta \ubs_f\|_\infty\leq 1$, which we have not included in our previous considerations. However, this constraint does not affect any of the previously discussed theory and (see the independence on input constraints in Theorem~\ref{thm:implicitPredictor2FCS}) and can be easily included in any of the computation methods considered in this section. Collection of persistently exciting I/O data is done by choosing random inputs adhering to $\ub(k)\in\Uc_k=\{-1, 0, 1\}^3$ and $\|\Delta\ub(k)\|_\infty\leq 1$ until the sufficient condition of persistent excitation according to the fundamental Lemma \cite{WILLEMS2005} is satisfied and a square (or wider) data matrix $\Dbc$ is obtained. To make the simulation more realistic and ensure that Assumption~\ref{assum:fullRank} is satisfied, we retroactively added measurement noise with a signal-to-noise ratio (SNR) of $40\text{dB}$ over the whole span of the data collection phase to each output channel. Qualitatively, one can expect the closed-loop performance to deteriorate for lower SNR, since both the SPC, and the implicit predictor become less accurate models of the true system dynamics.  The remaining parameters were chosen as $\lambda_a = 10^{3}, \Qb = \Ib, \Rb = 10^{-3}\Ib, N_p = 4$. 
We emphasize that the goal of this paper is not a comparison of performance between FCS-MPC and FCS-DPC. As we showed in Theorem~\ref{thm:implicitPredictor2FCS}, the predictive behavior of DPC is the same, regardless of whether FCS constraints are involved. Hence, this comparison of performance can be inferred from more general setups, like the ones considered in \cite{Dorfler2021}. Instead, our purpose is to  introduce an approach aimed at accelerating the computation times of FCS-DPC, while keeping its closed-loop behavior unchanged. Therefore, we instead compare the computation times (depending on $N_f$) for different methods of solving the FCS-DPC problem, namely: 
\begin{itemize}[align=left]
    \item[(ENUM)] Solving \eqref{eq:DPC} via enumeration of $\Uc$,
    \item[(MIQP)] \;\,Solving \eqref{eq:DPC} via MOSEK's \cite{mosek} MIQP solver,
    \item[(SDA)] \quad Solving \eqref{eq:FCS-DPC} via SDA.
\end{itemize}
Note that after a choice of $\ubs_f$, ENUM still needs to solve \eqref{eq:DPC} using a QP solver. For simplicity, we chose MATLAB's \cite{MATLAB} \texttt{quadprog} function with the \texttt{trust-region-reflective} algorithm, since the corresponding Hessian is generally positive semidefinite but positive definite on the nullspace of the equality constraints. With regard to SDA, we use conventional SDA with standard initialization as classified in \cite{Zafra2023}. We report statistics of the computation times in Fig.~\ref{tab:TimePlot}, which were recorded during closed-loop simulation over 800 time-steps each. Furthermore, we performed all simulations using the projection-based regularizer $h(\ab) = \|(\Ib-\Pib)\ab\|_2^2$ as the choice of regularizer did not significantly influence computation times. In our tests, SDA performs fastest among the three methods. However, it is also the method with the most computational variability (as discussed in \cite[Sec.~III.B]{Zafra2023}) and the most pre-processing involved, where some additional pre-processing could also benefit the other two methods. All in all, none of the computation times were able to reach the system example's sampling time of $25 \mathrm{\mu s}$ of the platform used. However, this example should be seen as a relative comparison of the three methods rather than a test of absolute performance.
\begin{rem}
    While we mentioned the reformulation \eqref{eq:FCS-DPC_indirect} to be equivalent in Remark~\ref{rem:directiVSindirect}, the actual computation revealed some minor (in terms of suboptimality) discrepancies of the optimal control sequences $\ubs_f^\ast$ obtained via the SDA approach. This difference is due to numerical errors involved in computing the inverses required for \eqref{eq:FCS-DPC}, especially with respect to the conditioning of, e.g., $\Qbc_{\text{reg}}$, where we observed a higher SNR to correlate with worse conditioning.
\end{rem}
\begin{figure}
    \centering
    \includegraphics[trim=1.8cm 9.9cm 2cm 8.4cm,clip=true, width=\linewidth]{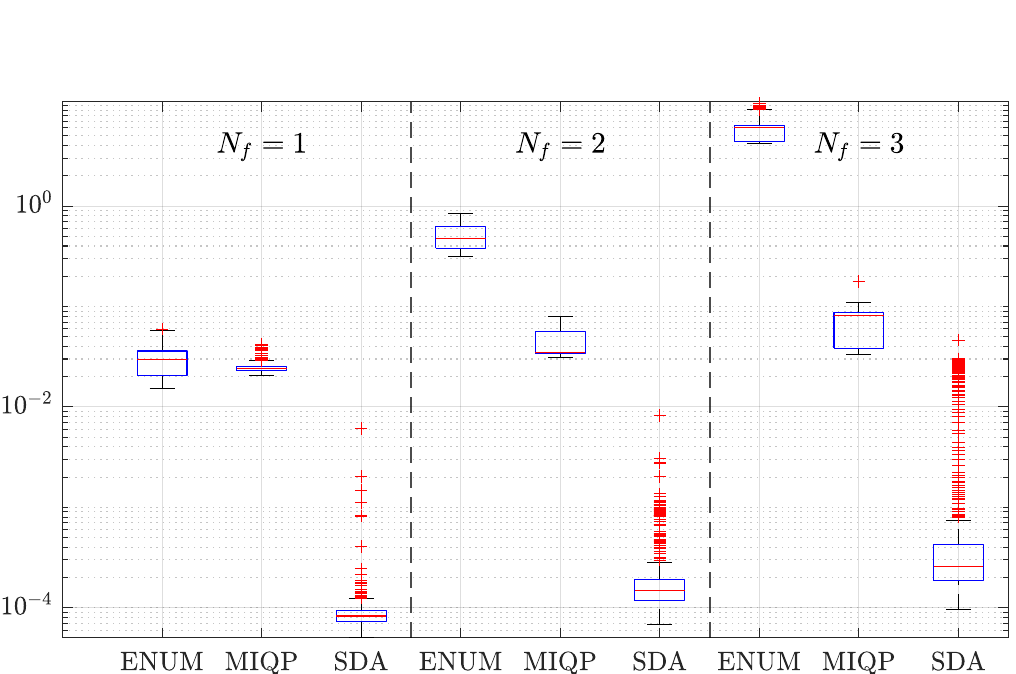}
        \caption{A boxplot showing computation times over 800 steps of closed-loop simulation. Crosses refer to outliers (all above the 75th percentile plus $1.5$ times the interquartile range).}
        \label{tab:TimePlot}
\end{figure}

\section{Conclusion and Outlook}\label{sec:Conclusion}

We showed how to reformulate the OCP associated with a standard FCS-DPC setup in order to enable application of the SDA popular in FCS-MPC. The reformulation was done by using the novel concept of implicit predictors, deriving an implicit predictor for the original OCP, and introducing it as an explicit equality constraint. A numerical example popular in FCS-MPC showed computation time improvements for using SDA in FCS-DPC compared to two readily available methods, i.e., an enumeration-based and a MIQP method.

Going forward, we aim to extend these ideas to be suitable for control of nonlinear systems with FCS constraints. Doing so might necessitate a thorough analysis of implicit predictors for different regularizers and other modifications commonly used in DPC setups for nonlinear systems.

% References

\end{document}